\documentclass[12pt,a4paper]{article}

\usepackage{lineno}  
\usepackage{graphicx}  
\usepackage{xspace} 
\usepackage{color}
\usepackage{colortbl}
\usepackage{amsmath} 
\usepackage{ifthen} 
\usepackage{subfigure}
\usepackage{bm}
\usepackage{ifthen} 

\newboolean{pdflatex}
\setboolean{pdflatex}{true} 
\usepackage{rotating} 
\newboolean{articletitles}
\setboolean{articletitles}{true} 

\newboolean{uprightparticles}
\setboolean{uprightparticles}{false} 

\usepackage{amssymb}
\usepackage{amsfonts}
\usepackage{upgreek} 

\usepackage{hyperref}    
\usepackage[all]{hypcap} 
\usepackage{float}
\usepackage{verbatim}
\usepackage{multicol}
 
\usepackage{xspace} 
\usepackage{upgreek}

\def\lhcb {\mbox{LHCb}\xspace}

\def\MagUp {\mbox{\em Mag\kern -0.05em Up}\xspace}

\ifthenelse{\boolean{uprightparticles}}%
{

 \def\Pmu         {\ensuremath{\upmu}\xspace}

 \def\Ppsi        {\ensuremath{\uppsi}\xspace}

 \def\PDelta      {\ensuremath{\Delta}\xspace}                 
 \def\PXi      {\ensuremath{\Xi}\xspace}                 
 \def\PLambda      {\ensuremath{\Lambda}\xspace}                 
 \def\PSigma      {\ensuremath{\Sigma}\xspace}                 
 \def\POmega      {\ensuremath{\Omega}\xspace}                 
 \def\PUpsilon      {\ensuremath{\Upsilon}\xspace}                 
 
 \def\PB      {\ensuremath{\mathrm{B}}\xspace}                 
                  
 \def\PD      {\ensuremath{\mathrm{D}}\xspace}

 \def\PJ      {\ensuremath{\mathrm{J}}\xspace}                 
 \def\PK      {\ensuremath{\mathrm{K}}\xspace}

 \def\Pb      {\ensuremath{\mathrm{b}}\xspace}                 
 \def\Pc      {\ensuremath{\mathrm{c}}\xspace}

 \def\Pi      {\ensuremath{\mathrm{i}}\xspace}

 \def\Ps      {\ensuremath{\mathrm{s}}\xspace}

}
{

 \def\Pmu         {\ensuremath{\mu}\xspace}

 \def\Ppsi        {\ensuremath{\psi}\xspace}                 
                  
 \mathchardef\PDelta="7101
 \mathchardef\PXi="7104
 \mathchardef\PLambda="7103
 \mathchardef\PSigma="7106
 \mathchardef\POmega="710A
 \mathchardef\PUpsilon="7107
                  
 \def\PB      {\ensuremath{B}\xspace}                 
                  
 \def\PD      {\ensuremath{D}\xspace}

 \def\PJ      {\ensuremath{J}\xspace}                 
 \def\PK      {\ensuremath{K}\xspace}

 \def\Pb      {\ensuremath{b}\xspace}                 
 \def\Pc      {\ensuremath{c}\xspace}

 \def\Pi      {\ensuremath{i}\xspace}

 \def\Ps      {\ensuremath{s}\xspace}

}

\makeatletter
\ifcase \@ptsize \relax
  \newcommand{\miniscule}{\@setfontsize\miniscule{4}{5}}
\or
  \newcommand{\miniscule}{\@setfontsize\miniscule{5}{6}}
\or
  \newcommand{\miniscule}{\@setfontsize\miniscule{5}{6}}
\fi
\makeatother

\DeclareRobustCommand{\optbar}[1]{\shortstack{{\miniscule (\rule[.5ex]{1.25em}{.18mm})}
  \\ [-.7ex] $#1$}}

\def\mumu       {{\ensuremath{\Pmu^+\Pmu^-}}\xspace}

\def\squark    {{\ensuremath{\Ps}}\xspace}

\def\cquark    {{\ensuremath{\Pc}}\xspace}

\def\bquark    {{\ensuremath{\Pb}}\xspace}

  \def\Kbar    {{\kern 0.2em\overline{\kern -0.2em \PK}{}}\xspace}

\def\KorKbar    {\kern 0.18em\optbar{\kern -0.18em K}{}\xspace}

  \def\Dbar    {{\kern 0.2em\overline{\kern -0.2em \PD}{}}\xspace}

\def\DorDbar    {\kern 0.18em\optbar{\kern -0.18em D}{}\xspace}

\def\B       {{\ensuremath{\PB}}\xspace}
\def\Bbar    {{\ensuremath{\kern 0.18em\overline{\kern -0.18em \PB}{}}}\xspace}

\def\BorBbar    {\kern 0.18em\optbar{\kern -0.18em B}{}\xspace}
\def\Bz      {{\ensuremath{\B^0}}\xspace}
\def\Bzb     {{\ensuremath{\Bbar{}^0}}\xspace}
\def\Bu      {{\ensuremath{\B^+}}\xspace}

\def\Bp      {{\ensuremath{\Bu}}\xspace}

\def\Bs      {{\ensuremath{\B^0_\squark}}\xspace}
\def\Bsb     {{\ensuremath{\Bbar{}^0_\squark}}\xspace}

\def\jpsi     {{\ensuremath{{\PJ\mskip -3mu/\mskip -2mu\Ppsi\mskip 2mu}}}\xspace}

  \def\Y#1S{\ensuremath{\PUpsilon{(#1S)}}\xspace}

\def\Lz          {{\ensuremath{\PLambda}}\xspace}
\def\Lbar        {{\ensuremath{\kern 0.1em\overline{\kern -0.1em\PLambda}}}\xspace}
\def\LorLbar    {\kern 0.18em\optbar{\kern -0.18em \PLambda}{}\xspace}

\def\Lb      {{\ensuremath{\Lz^0_\bquark}}\xspace}

\def\to                 {\ensuremath{\rightarrow}\xspace}

\def\AT#1     {\ensuremath{A_{\mathrm{T}}^{#1}}\xspace}           

\def\C#1      {\ensuremath{\mathcal{C}_{#1}}\xspace}                       
\def\Cp#1     {\ensuremath{\mathcal{C}_{#1}^{'}}\xspace}                    
\def\Ceff#1   {\ensuremath{\mathcal{C}_{#1}^{\mathrm{(eff)}}}\xspace}        
\def\Cpeff#1  {\ensuremath{\mathcal{C}_{#1}^{'\mathrm{(eff)}}}\xspace}       
\def\Ope#1    {\ensuremath{\mathcal{O}_{#1}}\xspace}                       
\def\Opep#1   {\ensuremath{\mathcal{O}_{#1}^{'}}\xspace}                    

\newcommand{\tev}{\ifthenelse{\boolean{inbibliography}}{\ensuremath{~T\kern -0.05em eV}\xspace}{\ensuremath{\mathrm{\,Te\kern -0.1em V}}}\xspace}
\newcommand{\gev}{\ensuremath{\mathrm{\,Ge\kern -0.1em V}}\xspace}
\newcommand{\mev}{\ensuremath{\mathrm{\,Me\kern -0.1em V}}\xspace}
\newcommand{\kev}{\ensuremath{\mathrm{\,ke\kern -0.1em V}}\xspace}
\newcommand{\ev}{\ensuremath{\mathrm{\,e\kern -0.1em V}}\xspace}
\newcommand{\gevc}{\ensuremath{{\mathrm{\,Ge\kern -0.1em V\!/}c}}\xspace}
\newcommand{\mevc}{\ensuremath{{\mathrm{\,Me\kern -0.1em V\!/}c}}\xspace}
\newcommand{\gevcc}{\ensuremath{{\mathrm{\,Ge\kern -0.1em V\!/}c^2}}\xspace}
\newcommand{\gevgevcccc}{\ensuremath{{\mathrm{\,Ge\kern -0.1em V^2\!/}c^4}}\xspace}
\newcommand{\mevcc}{\ensuremath{{\mathrm{\,Me\kern -0.1em V\!/}c^2}}\xspace}

\def\mum  {\ensuremath{{\,\upmu\mathrm{m}}}\xspace}

\def\invfb   {\ensuremath{\mbox{\,fb}^{-1}}\xspace}

\def\ps   {\ensuremath{{\mathrm{ \,ps}}}\xspace}

\newcommand{\chisq}{\ensuremath{\chi^2}\xspace}

\newcommand{\chisqip}{\ensuremath{\chi^2_{\text{IP}}}\xspace}

\def\gsim{{~\raise.15em\hbox{$>$}\kern-.85em
          \lower.35em\hbox{$\sim$}~}\xspace}
\def\lsim{{~\raise.15em\hbox{$<$}\kern-.85em
          \lower.35em\hbox{$\sim$}~}\xspace}

\def\ptot       {\mbox{$p$}\xspace}
\def\pt         {\mbox{$p_{\mathrm{ T}}$}\xspace}

\def\evtgen     {\mbox{\textsc{EvtGen}}\xspace}

\def\geant      {\mbox{\textsc{Geant4}}\xspace}

\def\photos     {\mbox{\textsc{Photos}}\xspace}

\def\pythia     {\mbox{\textsc{Pythia}}\xspace}

\def\tell1  {TELL1\xspace}
\def\ukl1   {UKL1\xspace}

\newcommand{\ie}{\mbox{\itshape i.e.}\xspace}

\def\PTheta {\ensuremath{\Theta}\xspace}

\newcommand*\patchAmsMathEnvironmentForLineno[1]{%
\expandafter\let\csname old#1\expandafter\endcsname\csname #1\endcsname
\expandafter\let\csname oldend#1\expandafter\endcsname\csname
end#1\endcsname
 \renewenvironment{#1}%
   {\linenomath\csname old#1\endcsname}%
   {\csname oldend#1\endcsname\endlinenomath}%
}
\newcommand*\patchBothAmsMathEnvironmentsForLineno[1]{%
  \patchAmsMathEnvironmentForLineno{#1}%
  \patchAmsMathEnvironmentForLineno{#1*}%
}
\AtBeginDocument{%
\patchBothAmsMathEnvironmentsForLineno{equation}%
\patchBothAmsMathEnvironmentsForLineno{align}%
\patchBothAmsMathEnvironmentsForLineno{flalign}%
\patchBothAmsMathEnvironmentsForLineno{alignat}%
\patchBothAmsMathEnvironmentsForLineno{gather}%
\patchBothAmsMathEnvironmentsForLineno{multline}%
\patchBothAmsMathEnvironmentsForLineno{eqnarray}%
}

\usepackage[bottom]{footmisc}
\usepackage{cite}
\usepackage{mciteplus}
\usepackage{afterpage}
\usepackage{epsfig,accents}
\usepackage[toc,page]{appendix}
\begin{document}






\renewcommand{\thefootnote}{\fnsymbol{footnote}}

\setcounter{footnote}{1}

\begin{titlepage}

\vspace*{-1.5cm}

\hspace*{-0.5cm}
\begin{tabular*}{\linewidth}{lc@{\extracolsep{\fill}}r}
\ifthenelse{\boolean{pdflatex}}
{\vspace*{-1.7cm}\mbox{\!\!\!\includegraphics[width=.14\textwidth]{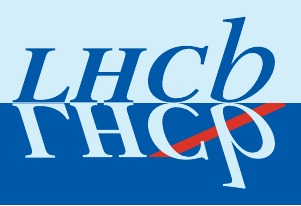}} & &}%
{\vspace*{-1.2cm}\mbox{\!\!\!\includegraphics[width=.12\textwidth]{lhcb-logo.eps}} & &}
 \\
 & & CERN-EP-2017-329\\
 & &LHCb-PAPER-2017-043\\  
 & & \today \\ 
 & & \\
\hline
\end{tabular*}

\vspace*{4.0cm}

{\bf\huge
\begin{center}
{\boldmath A search for weakly decaying $b$-flavored pentaquarks}
\end{center}
}

\vspace*{2.0cm}

\begin{center}
The LHCb collaboration\footnote{Authors are listed at the end of this paper.}
\end{center}

\vspace{\fill}

\vspace{\fill}

\begin{abstract}
  \noindent
Investigations of the existence of pentaquark states containing a single $b$ (anti)quark decaying weakly into four specific final states  $\jpsi K^+\pi^- p$, $\jpsi K^- \pi^- p$, $\jpsi K^- \pi^+ p$, and $\jpsi \phi(1020) p$ are reported.  The data sample corresponds to an integrated luminosity of 3.0~\!\!\invfb in 7 and 8~\!TeV $pp$ collisions acquired with the LHCb detector. Signals are not observed and upper limits are set on the product of the production cross section times branching fraction with respect to that of the  \Lb.

\end{abstract}

\vspace*{2.0cm}
\vspace{\fill}

\begin{center}
Submitted to Physical Review D 
\end{center}

\vspace{\fill}

{\footnotesize 
\centerline{\copyright~CERN on behalf of the \lhcb collaboration, licence \href{http://creativecommons.org/licenses/by/4.0/}{CC-BY-4.0}.}}
\vspace*{2mm}

\end{titlepage}

\renewcommand{\thefootnote}{\arabic{footnote}}
\setcounter{footnote}{0}

\pagestyle{empty}  



\setcounter{page}{2}
\mbox{~}

\cleardoublepage



\pagestyle{plain} 
\setcounter{page}{1}
\pagenumbering{arabic}


%
\newpage
\section{Introduction}

The observation of charmonium pentaquark states with quark content $c\overline{c}uud$, by the LHCb \cite{Aaij:2015tga} collaboration in $\Lb \to \jpsi K^- p $ decays, raises many questions including: What is the internal structure of these pentaquarks? Do other pentaquark states exist?  Are they molecular or tightly bound?  In this analysis, we search for pentaquarks that contain a single $b$ (anti)quark, that decay via the weak interaction. The Skyrme model \cite{Skyrme:1961vq} has been used to predict that the heavier the constituent quarks, the more tightly bound the pentaquark state \cite{Klebanov,Stewart:2004pd,Leibovich:2003tw,Oh:1994np}. This motivates our search for pentaquarks containing a $b$ (anti)quark. No existing searches for weakly decaying pentaquarks containing a $b$ (anti)quark have been published.

\begin{figure}[b]
\centering
\includegraphics[width=1.0\textwidth]{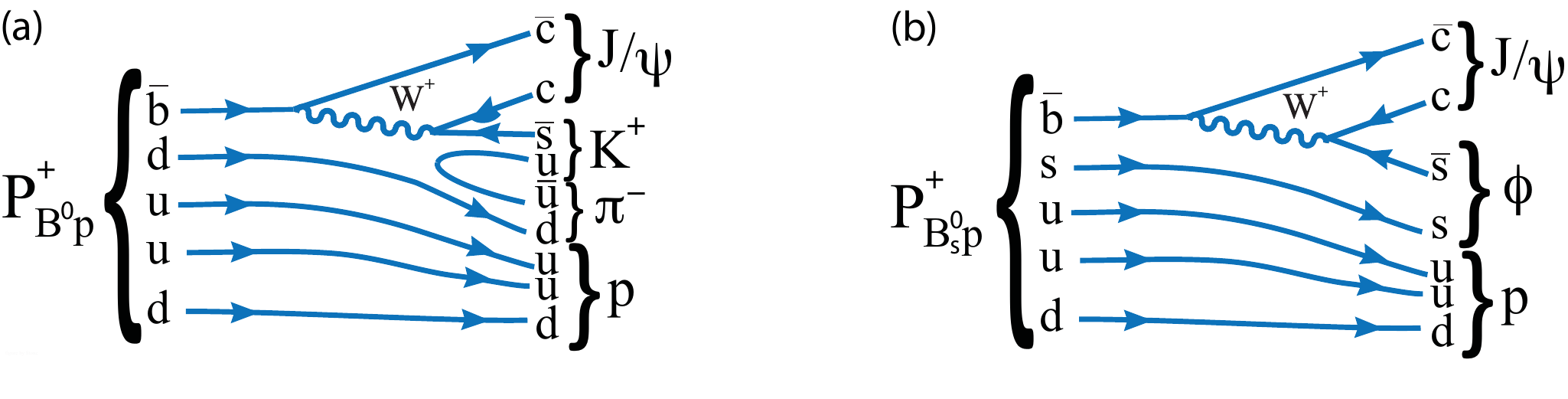}
\vskip -0.25cm
\caption{\small Leading-order diagrams for pentaquark decay modes into (a) $\jpsi K^+ \pi^- p$ or (b) $\jpsi \phi p$ final states.}
\label{Feyn0}
\end{figure}

Consider the possible pentaquark states $\overline{b}duud$, $b\overline{u}udd$, $b\overline{d}uud$ and $\overline{b}suud$. We label these states as $P_{\Bz p}^+$, $P_{\Lb \pi^-}^-$, $P_{\Lb \pi^+}^+$ and $P_{\Bs p}^+$, respectively, where the subscript indicates the final states the pentaquark would predominantly decay into if it had sufficient mass to decay strongly into those states. While there are many possible decay modes of these states, we focus  on modes containing a \jpsi meson in the final state because these candidates generally have relatively large efficiencies and reduced backgrounds in the LHCb experiment. The Feynman diagrams for the decay of the $P_{\Bz p}^+$ and $P_{\Bs p}^+$ states are shown in Fig.~\ref{Feyn0}. The corresponding diagrams for the decay of $P_{\Lb \pi^-}^-$ and $P_{\Lb \pi^+}^+$ are similar to that shown in Fig.~\ref{Feyn0}(a), with the decay of the state being driven by the $b \to c\overline{c}s$ transition. We reconstruct the $\phi(1020)$ meson\footnote{Hereafter $\phi$ refers to the $\phi(1020)$ meson.} in the $K^+K^-$ decay mode. We note that the $P^+_{\Bz p}$ pentaquark might have some decays inhibited by Bose statistics if its structure is based on two identical $ud$ diquarks, \ie $\overline{b}(ud)(ud)$. Although the $P^+_{\Bs p}$ state is expected to be produced at a smaller rate on the grounds that $\Bs$ production in the LHCb experiment acceptance is only about 13\% of the rate of the sum of $\Bp$ and $\Bz$ production \cite{Aaij:2011jp}, it would not have two identical diquarks, and hence none of its decays would suffer from spin-statistics suppression.

Table~\ref{tab:modes} lists all of the pentaquarks we search for along with their respective weak decay modes.\footnote{Unless explicitly stated, mention of a particular mode implies the use of the charge-conjugated mode as well.}
\begin{table}[t]
\centering
\caption{\small Quark content of the $b$-flavored pentaquarks and their weak decay modes explored here. We consider only the quark decay process  $b \to c\overline{c}s$.  The lower and upper bounds of the mass region searched are also given. (In this paper we use natural units where $\hbar=c=1$.)}
\vspace{0.2cm}
\begin{tabular}{rclcc}
\hline
~~~~Mode&Quark content &~~ Decay mode& Search window \\
\hline
\textbf{I}&$\overline{b}duud$  & $P_{\Bz p}^+~\to \jpsi K^+\pi^-p$ &4668--6220~MeV \\ 
\textbf{II}&$b\overline{u}udd$   &$P_{\Lb \pi^-}^- \! \to \jpsi K^- \pi^- p$&4668--5760~MeV\\   
\textbf{III}&$b\overline{d}uud$  & $P_{\Lb \pi^+}^+  \! \to \jpsi K^- \pi^+ p$&4668--5760~MeV\\         
\textbf{IV}&$\overline{b}suud$ & $P_{B_s^0 p}^+  ~\to \jpsi\phi p$&5055--6305~MeV\\                                                 
                                    
\hline 
\end{tabular}
\label{tab:modes}
\end{table}
It is possible for these pentaquarks ($P_B$) to decay either strongly or weakly depending on their masses. The threshold mass for strong decay for $P_{\Bz p}^+$ would be $m(\Bz)+m(p)$, for $P_{\Lb \pi^-}^-$ $m(\Lb)+m(\pi^-)$, for $P_{\Lb \pi^+}^+$ 
$m( \Lb)+m(\pi^+$) and for $P_{\Bs p}^+$ $m(B_s^0)+m(p$). Therefore,  we define our signal search windows to be below these thresholds. Note that a fifth state, the $b\overline{s}uud$ pentaquark ($P^+_{\Bsb p}$) could also decay into $\jpsi \phi p$, and thus is implicitly included in our searches. Should a signal be detected for mode \textbf{IV}, we would need to examine noncharmonium modes to distinguish between the possibilities. 

\section{Detector description and data samples}
\label{sec:det_data}
The \lhcb detector~\cite{Alves:2008zz,Aaij:2014jba} is a single-arm forward
spectrometer covering the \mbox{pseudorapidity} range $2<\eta<5$,
designed for the study of particles containing \bquark or \cquark
quarks. The detector includes a high-precision tracking system
consisting of a silicon-strip vertex detector surrounding the $pp$
interaction region, a large-area silicon-strip detector located
upstream of a dipole magnet with a bending power of about
$4{\mathrm{\,Tm}}$, and three stations of silicon-strip detectors and straw
drift tubes placed downstream of the magnet.
The tracking system provides a measurement of momentum, \ptot, of charged particles with
a relative uncertainty that varies from 0.5\% at low momentum to 1.0\% at 200\gev.
The minimum distance of a track to a primary $pp$ interaction vertex (PV), the impact parameter (IP), 
is measured with a resolution of $(15+29/\pt)\mum$,
where \pt is the component of the momentum transverse to the beam, in\,\gev.
Different types of charged hadrons are distinguished using information
from two ring-imaging Cherenkov detectors (RICH). 
Photons, electrons and hadrons are identified by a calorimeter system consisting of
scintillating-pad and preshower detectors, an electromagnetic
calorimeter and a hadronic calorimeter. Muons are identified by a
system composed of alternating layers of iron and multiwire
proportional chambers.

The online event selection is performed by a trigger, 
which consists of a hardware stage, based on information from the calorimeter and muon
systems, followed by a software stage, which applies a full event
reconstruction. The subsequent software trigger is composed of two stages, the first of which performs a partial reconstruction and requires either a pair of well-reconstructed, oppositely charged muons having an invariant mass above 2.7\gev, or a single well-reconstructed muon with high \pt and large IP. The second stage of the software trigger applies a full event reconstruction and, for this analysis, requires two opposite-sign muons to form a good-quality vertex that is well separated from all of the PVs, and to have an invariant mass within $\pm120$\mev of the known \jpsi mass~\cite{PDG}. The data sample corresponds to 1.0~\!\!\invfb of integrated luminosity collected with the LHCb detector in 7~\!TeV $pp$ collisions and 2.0~\!\!\invfb in 8~\!TeV collisions.

Simulated events are generated in the LHCb acceptance using \pythia
  \cite{Sjostrand:2006za,*Sjostrand:2007gs}, with a special LHCb parameter tune \cite{LHCb-PROC-2010-056}. Pentaquark candidate ($P_B$) decays are generated uniformly in phase space. Decays of other hadronic particles are described by \evtgen~\cite{Lange:2001uf}, in which final-state radiation is generated using \photos~\cite{Golonka:2005pn}. The interaction of the generated particles with the detector, and its response, are implemented using the \geant toolkit~\cite{Allison:2006ve, *Agostinelli:2002hh} as described in
Ref.~\cite{LHCb-PROC-2011-006}. The lifetime of the simulated pentaquarks is set to 1.5\ps, consistent with that of most weakly decaying $b$ hadrons \cite{PDG}.

\section{\boldmath Event selection and $b$-hadron reconstruction}
\label{sec:selection}

A pentaquark candidate is reconstructed by combining a $\jpsi \to \mu^+\mu^-$ candidate with a proton, kaon, and pion (or kaon for mode \textbf{IV}). Our analysis strategy consists of a preselection based on loose particle identification (PID) and the kinematics of the decay, followed by a more sophisticated multivariate selection (MVA) classifier based on a  Boosted Decision Tree (BDT)~\cite{Breiman}, which uses multiple input variables, accounts for the correlations and outputs a single discriminant. In order to avoid bias, the data in the signal search regions were not examined (blinded) until all the selection requirements were decided.

In the preselection, the \jpsi candidates are formed from two oppositely charged particles with $\pt$ greater than 500$\mev$, identified as muons and consistent with originating from a common vertex but inconsistent with originating from any PV. The invariant mass of the $\mumu$ pair is required to be within $[-48, +43]~\!$MeV of the known \jpsi mass~\cite{PDG}, corresponding to a window of about $\pm3$ times the mass resolution. The asymmetry in the mass window is due to the radiative tail. Pion, kaon, and proton candidates are required to be positively identified in the RICH detector, but with loose requirements as the MVA includes particle identification criteria. Kaon and proton candidates are required to have momenta greater than 5~GeV and 10~GeV, respectively, to avoid regions with suboptimal particle identification.
Each track must have an IP $\chi^2$ greater 9 than with respect to the closest PV, must have $\pt$ greater than 250\mev, and the scalar sum of the tracks $\pt$ is required to be larger than 900\mev. 
 All of the tracks forming the pentaquark state are required to form a good vertex and have a significant detachment from the PV. We also require that the cosine of the angle between the vector from the PV to the $P_B$ candidate vertex ($\vec{V}_{\!\!PV-P_B}$) and the $P_B$ candidate momentum vector ($\vec{p}_{P_B}$) be greater than $0.999$. The invariant mass of the pentaquark states is calculated by constraining the invariant mass of the dimuon pair to the known $\jpsi$ mass, the muon tracks to originate from the $\jpsi$ vertex and the vector sum of the momenta of the final state particles to point back to the PV.
 
We measure the product of the production cross section and branching fraction of these pentaquark states and normalize it to the analogous measurement \cite{Aaij:2015Lb} by the LHCb collaboration for the $\Lb$ baryon in the $\Lb\to\jpsi K^- p$ decay. To this end, we impose the same kinematic requirements on the $P_B$ candidate as applied to the $\Lb$ candidates in that analysis, namely  $p_T < 20$ GeV and $2.0 < y < 4.5$, where $y = \frac{1}{2} \ln\left(\frac{E + p_z}{E - p_z}\right)$ is the rapidity, $E$ the energy and $p_z$ the component of the momentum along the beam direction. After these preselections, the product of trigger and reconstruction efficiencies is around 2\% for all the modes.

\section{Selection optimization by a multivariate classifier}
 
\begin{figure}[t]
\centering
\includegraphics[width= .67\textwidth]{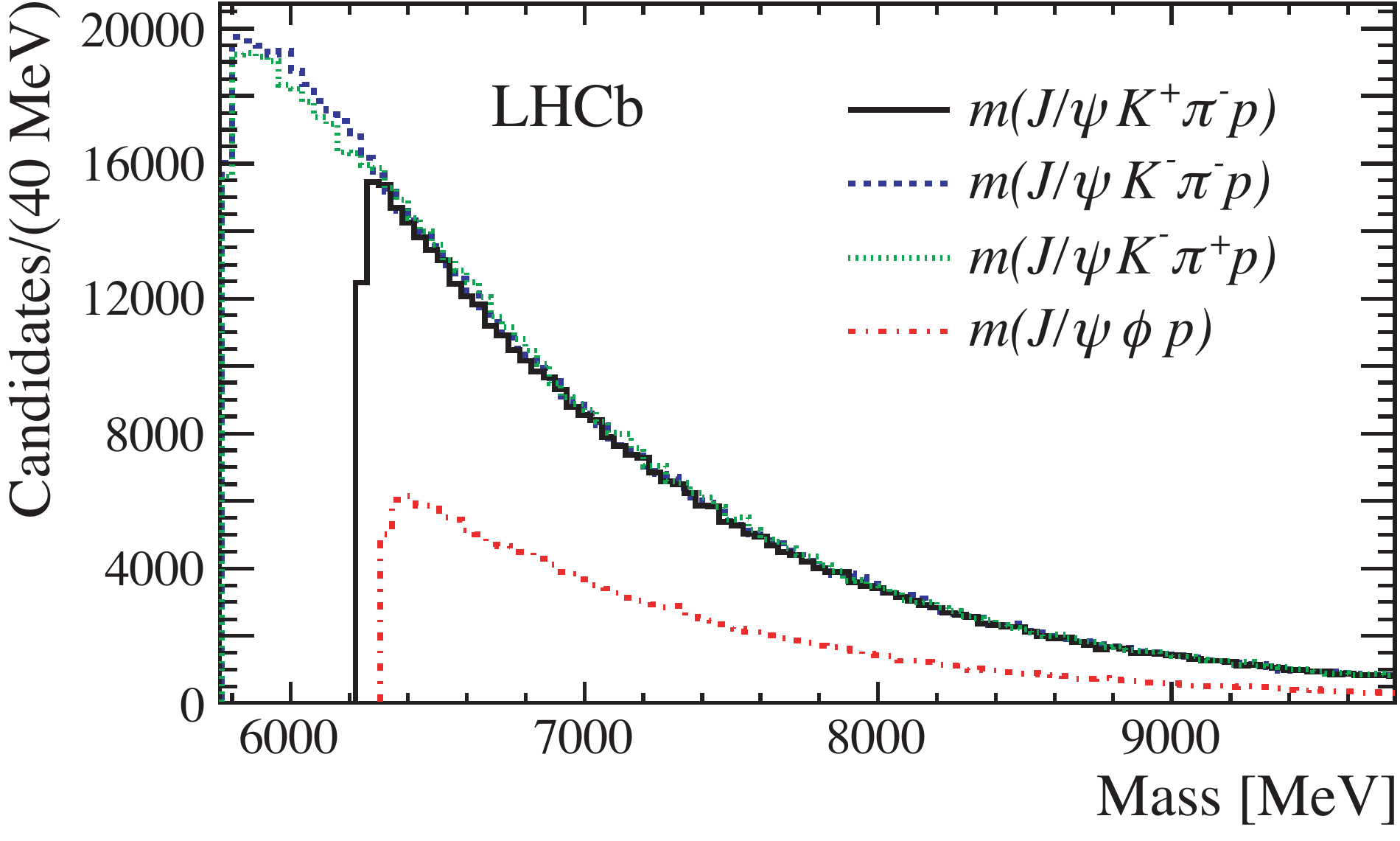}
\vskip -2.0mm
\caption{\small Invariant mass distributions above the decay mass thresholds for the indicated modes.}
\label{JpsiKPiP_Bkg_Mass}
\end{figure}

The MVA classifier is trained using the simulated signal samples described at the end of Section \ref{sec:det_data} and a background sample of candidates in data with invariant masses within 0.5\gev above the strong-decay threshold in each final state (see Fig.~\ref{JpsiKPiP_Bkg_Mass}).
We use $3 \times 10^6 $ $P_{\Bz p}^+\to (\jpsi \to \mu^+ \mu^-) K^+\pi^- p$ simulated events for modes \textbf{I}, \textbf{II} and \textbf{III}, with the  $P_{\Bz p}^+$ mass set to 5750 MeV, and $3 \times 10^6 $ $P_{\Bs p}^+\to(\jpsi \to \mu^+ \mu^-) (\phi \to K^+ K^-) p$ simulated events for mode \textbf{IV}, with the $P_{\Bs p}^+$ mass set to 5835 MeV. The dependence of the selection efficiency as a function of mass is accounted for in Section \ref{sec:results}.

\begin{figure}[b]
\centering
\includegraphics[width=1.0\textwidth]{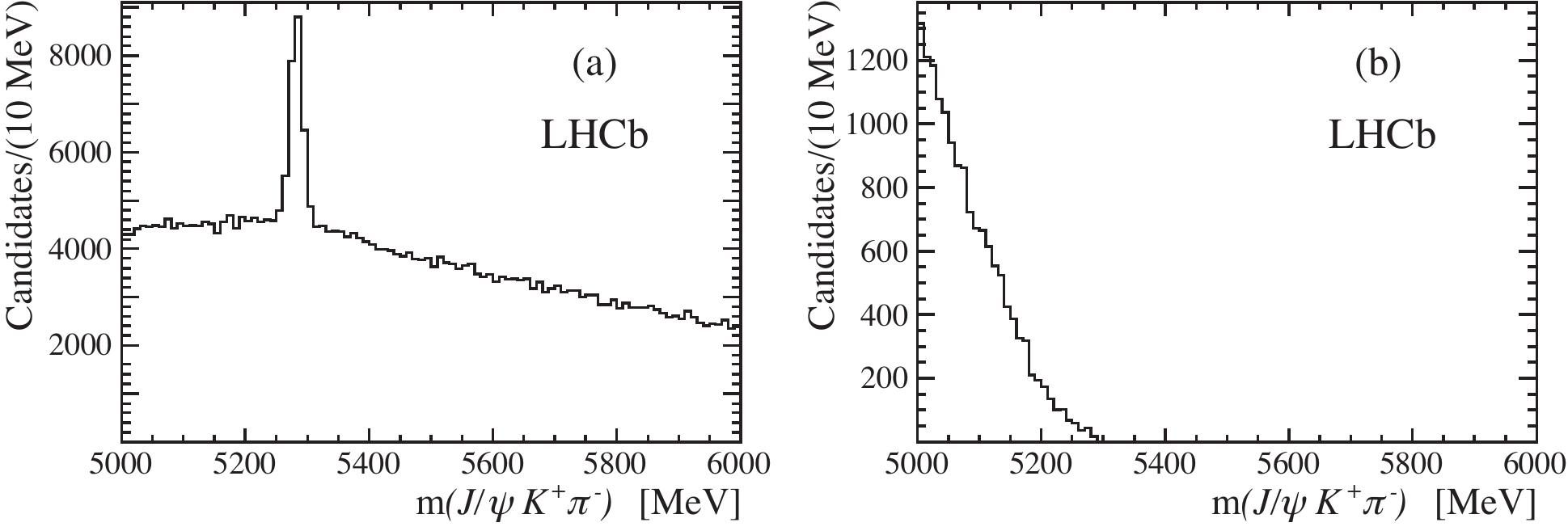}
\vskip -2mm
\caption{\small  For the ${P_{\Bz p}^+\to \jpsi K^+\pi^- p}$ decay search (mode \textbf{I}), the invariant mass of  $ \jpsi K^+ \pi^-$ combinations in the (a) region above threshold  and in the (b) signal region. The peak in the sideband region results from \Bz decays.}
\label{fig:vetoes}
\end{figure}

The training samples needed to model the backgrounds in the signal regions must represent the actual backgrounds as closely as possible.  Contamination in the background samples can occur from fully reconstructed weakly decaying $b$-hadrons that are combined with random particles. 
 In mode \textbf{I}, we find contributions from  $\Bz \to \jpsi K^+ \pi^-$ decays and $B_s^0 \to \jpsi K^+ K^-$ decays where one of the kaons is misidentified as a pion; then a random additional proton results in contamination in the background sample. In modes \textbf{II} and \textbf{III}, along with the $\Bz$ and $B_s^0$ contaminations, a $\Lb\to\jpsi K^- p$ decay can be paired with a random pion. In mode \textbf{IV}, only the $\Bz$ and $B_s^0$ contaminations are seen. 
These mistaken identification contributions in the background sample are found by looking at the invariant mass distributions obtained by switching one or more final-state particles to another mass hypothesis. If this produces a peak in the mass distribution at the mass of a known particle, we apply a veto in the background training sample eliminating all candidates within $\pm 12~\!$MeV of the peaks, approximately $\pm 1.6~\sigma$. No such peaks are seen in the signal region, after switching the mass hypotheses, for any of the modes. As an example, we show fully reconstructed decays in the background and signal regions for mode \textbf{I} in  Fig.~\ref{fig:vetoes}. 

The input variables used to train the classifier for modes \textbf{I}, \textbf{II}, and \textbf{III} are the same. We use the difference in the logarithm of the likelihood for two different particle hypotheses (DLL). They are the DLL($\mu-\pi$) for the two muons,  DLL($K-\pi$) and DLL($K-p$)  for the kaon,  DLL($p-\pi$) and DLL($p-K$) for the proton, and DLL($\pi-K$) for the pion. 
Also used is the logarithm of \chisqip, defined as the difference in \chisq of a given PV reconstructed with and
without the considered $K$, $\pi$, and $p$ tracks, and the  $\chi^2$ of the $P_B$ to be consistent with originating from the PV. Other variables are the logarithm of the cosine of the angle of  $\vec{p}_{P_B}$ with $\vec{V}_{\!PV-P_B}$, \ the flight distance of $P_{B}$, the scalar sum \pt of the $K$, $\pi$ and $p$ tracks, the $\chi^2$/ndof of the fit of all the decay tracks to the $P_B$ vertex, and of the two muon tracks to the \jpsi vertex with constraints that fix the dimuon invariant mass to the \jpsi mass and force the $P_{B}$ candidate to point back to the PV, where ndof indicates the number of degrees of freedom.  The input variables used to train the classifier for mode \textbf{IV} are similar, but with two kaons instead of a kaon and a pion.

\begin{figure}[t]
\centering
\includegraphics[width= 0.98\textwidth]{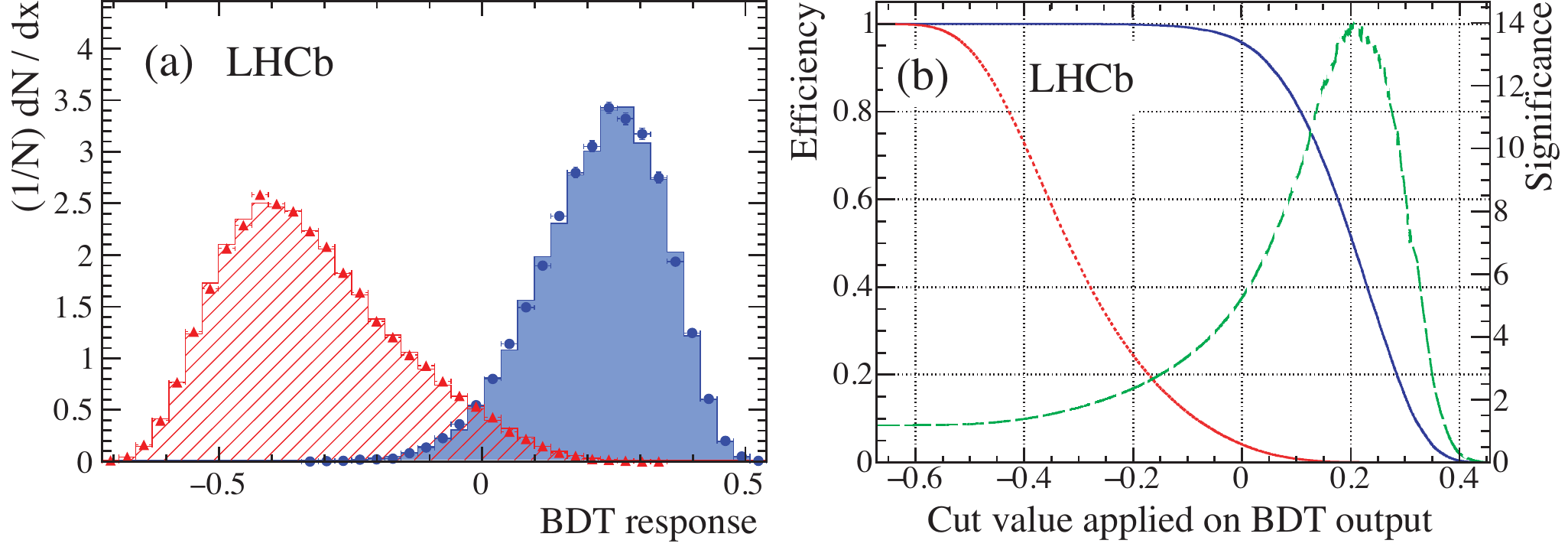}
\vskip -2mm
\caption{\small (a) Outputs of the BDT classifier for the $\jpsi K^+\pi^- p$ final state. The circles (blue) show the signal training sample, and the triangles (red) show the background training sample, while the shaded (blue) histogram shows the signal test sample, and the diagonal (red) line-shaded histogram the background test sample. (b) Efficiencies of signal, solid (blue) curve, and background, dotted (red) curve, and the value of the $S/(\sqrt{B}+1.5)$, dashed (green) curve, called ``significance," as a function of the BDT output. }
\label{fig:BDT1}
\end{figure}

Two important attributes of multivariate classifiers are signal efficiency and background rejection, both of which we wish to maximize. Using the input variables and training samples described earlier, we compared the performances of some common classifiers, including Boosted Decision Trees (BDT), Gradient Boosted Decision Trees, Linear Discriminant, and Likelihood Estimators \cite{Hocker:2007ht}. We base our MVA selection on the BDT algorithm. Once the BDT classifier is trained, it is evaluated by applying it to a separate testing sample (which is disjoint from the data sample used to train the classifier). The classifier assigns a response (called the BDT output) valued between --1 and 1 to the events, with background events tending toward low values and signal events to high values. These can be seen in Fig.~\ref{fig:BDT1}(a) for mode \textbf{I}. The BDT outputs for other modes look very similar.

Discrimination between signal candidates, $S$, and background, $B$, is accomplished by choosing a BDT value that maximizes
the metric $\frac{S}{a/2+\sqrt{B}}$, where $a$ is the significance of the signal sought,  which has the advantage of being  independent of the signal cross section \cite{Punzi:2003bu}. We choose $a$ to be 3 for all modes, based on the assumption that we are in a situation of looking for a small signal in the midst of larger backgrounds. The variation of the signal and background efficiencies and the metric's value with the BDT output is shown in Fig.~\ref{fig:BDT1}(b) for mode {\textbf{I}}. This variation of efficiencies and the metric with respect to the BDT value is similar for the other modes. After optimization, the BDT signal efficiency varies from 42.9\% to 71.4\% depending on the decay mode.

One cause of concern is reflections where the particle identification fails leading to the inclusion of other well-known final states. These are eliminated with a small loss of efficiency by removing candidate combinations within $\pm12~\!$MeV of the appropriate $b$-hadron mass. A list of these reflections in the particular modes of interest is given in Table~\ref{tab:reflect}.

\begin{table}[b]
\centering
\caption{\small Decay modes that are vetoed for each pentaquark candidate mode and the specific particle misidentification that causes the reflection.}
\vspace{0.2cm}
\begin{tabular}{lll}
\hline
~~~~Search mode & ~~~~Reflection & Particle misidentification\\
\hline
$P^+_{B^0 p}~\to \jpsi K^+\pi^- p$    & $B^+\to\jpsi K^+\pi^-\pi^+$ &~~~$\pi^+$ to  $p$ \\  
&$B^+\to\jpsi \pi^+\pi^-K^+$ & ~~~$\pi^+$ to $K^+$ and $K^+$ to $p$\\
$P^-_{\Lb\pi^-}\!\to \jpsi K^-\pi^- p$    & $B^-\to\jpsi K^-\pi^-\pi^+$ & ~~~$\pi^+$ to $p$\\        
&$B^- \to \jpsi (\phi \to K^- K^+)\pi^- $ & ~~~$K^+$ to $p$\\                              
 $P^+_{\Lb\pi^+}\!\to \jpsi K^-\pi^+ p$    & $B^+ \to \jpsi (\phi \to K^- K^+)\pi^+$  & ~~~$K^+$ to $p$ \\   
$P^+_{\Bs p}~\to \jpsi  \phi p$    & $B^+\to\jpsi \phi K^+$ &~~ $K^+$  to $p$ \\                                     
\hline 
\end{tabular}
\label{tab:reflect}
\end{table}

\section{Results}
\label{sec:results}
\begin{figure}[t]
\centering
\includegraphics[width=1.0\textwidth]{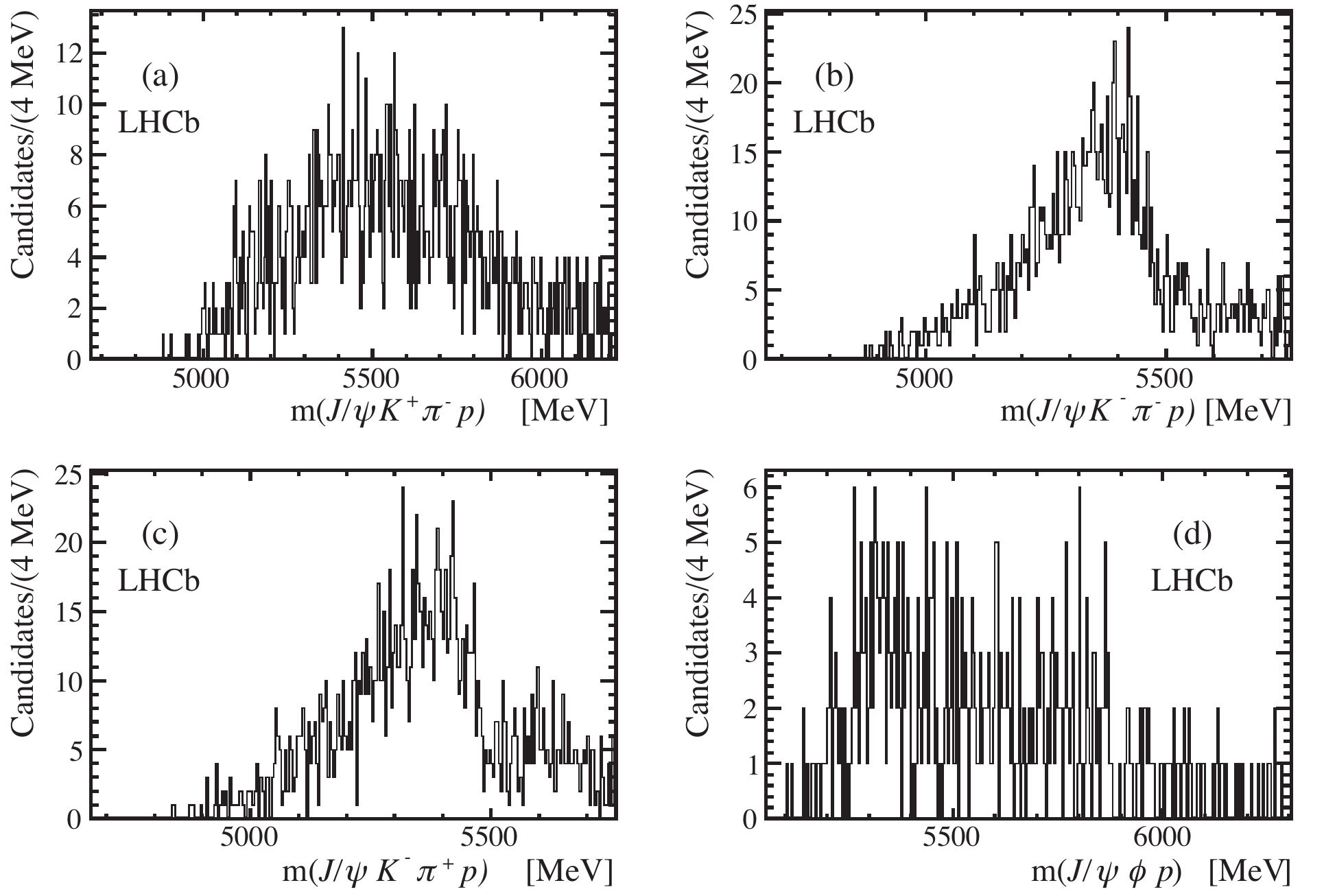}
\vskip -0.1cm
\caption{\small Reconstructed mass distributions after the BDT selection for the (a) $\jpsi K^+\pi^- p$, (b) $\jpsi K^- \pi^- p,$ (c) $\jpsi K^-\pi^+ p$, and (d) $\jpsi \phi p$ final states.}
\label{finalmass}
\end{figure}

After the selections were decided upon, the analysis was unblinded. 
A search is conducted by scanning the $P_B$ invariant mass distributions in the four final states shown in Fig.~\ref{finalmass}.
The step size used in these scans is 4.0\mev, corresponding to about half the invariant mass resolution. 
No signal is observed with the expected width of approximately 7.5 MeV. The $P_B$ mass resolution seen in the simulated samples is 6.0~MeV for modes \textbf{I}, \textbf{II}, \textbf{III}, and 5.2~MeV for mode \textbf{IV} which, as expected, is similar to the 7.5~MeV width seen in data for the $\Lb$ baryon in the $(\jpsi \to \mu^+ \mu^-) K^- p$ final state, when the two muons are constrained to the $\jpsi$ mass. In order to obtain conservative results, we set upper limits based on the wider 7.5~MeV signal width. 

At each $P_B$ scan mass value $m_{P_B}$, the signal region is a $\pm2\sigma(m_{P_B})$ window around $m_{P_B}$, while the background is estimated by interpolating the yields in the sidebands starting at $3\sigma(m_{P_B})$ from $m_{P_B}$  and extending to $5\sigma(m_{P_B})$, both below and above $m_{P_B}$  following Ref.~\cite{WILLIAMS2015}.
The statistical test at each mass is based on the profile likelihood ratio of Poisson-process hypotheses with and without a signal contribution, where the uncertainty on the background interpolation is modeled as purely Poisson (see Ref.~\cite{WILLIAMS2015} for details).
No significant excess of signal candidates is observed over the expected background. The upper limits are set on the signal yields using the profile likelihood technique, in which systematic uncertainties are handled by including additional Gaussian terms in the likelihood.

In the absence of a significant signal, we set upper limits in each $P_B$ candidate mass interval 
on the ratio 
\begin{equation}
\label{eq:R}
R=\frac{\sigma(pp\to P_B X)\cdot {\cal{B}}(P_B\to \jpsi X)}{\sigma(pp\to \Lb X)\cdot {\cal{B}}(\Lb\to\jpsi K^- p)}~,
\end{equation}
where we use the $\Lb \to \jpsi K^- p$ channel for normalization. The product of the production cross section and branching fraction of this channel has been measured by the LHCb collaboration \cite{Aaij:2015Lb} to be
\begin{equation}\label{eq:norm}
\begin{aligned}
\sigma(\Lb, \sqrt{s} = 7 \, {\rm TeV}) \cdot {\cal{B}}(\Lb \to \jpsi K^- p) = 6.12 \pm 0.10  \pm 0.25 \,{\rm nb},\\
\sigma(\Lb, \sqrt{s} = 8 \, {\rm TeV}) \cdot {\cal{B}}(\Lb \to \jpsi K^- p) = 7.51 \pm 0.08 \pm 0.31 \,{\rm nb},
\end{aligned}
\end{equation}
where the uncertainties are statistical and systematic, respectively. The systematic uncertainties include those on the luminosity and detection efficiencies that partially cancel, lowering the effective systematic uncertainty on the normalization. These measurements are averaged, taking into account the different luminosities at the two energies, to produce the overall normalization factor of  $NF=7.03\pm 0.06\pm 0.17$~nb.

Simulations have been generated at four different $P_B$ masses for each decay mode.  The total selection efficiency varies from 0.45\% to 1.4\% depending on mass and decay mode.  The mass dependence of the efficiencies is parametrized by a second-order polynomial, for each decay mode, and incorporated into the upper limit calculation. The dominant source of uncertainty on the efficiency is systematic, and arises from the calibration applied to the particle identification as calculated by the simulation. This absolute efficiency uncertainty varies from 0.02\% to 0.17\% depending on the decay mode. The statistical uncertainties on the efficiency are negligible. Note that we are taking the $P_B$ lifetime as 1.5~ps, and all simulated efficiencies assume that the $P_B$ decays are given by phase space.

For modes \textbf{I},  \textbf{II}, and \textbf{III}, the upper limits on $S$ are normalized to obtain the upper limits on $R$ according to

\begin{equation}
\label{eq:ul1}
{\rm UL}  (R) = \frac{{\rm UL} (S)}{{\cal{L}} \cdot {\cal{B}}(\jpsi \to \mu^+ \mu^-) \cdot   NF}~ ,
\end{equation}
where ${\rm UL} (S)$ is the efficiency corrected upper limit on $S$ in each particular mass bin, ${\cal{L}}$ is the integrated  luminosity and ${\cal{B}}(\jpsi \to \mu^+ \mu^-)$ is the branching fraction for the $\jpsi \to \mu^+ \mu^-$ decay. For mode \textbf{IV}, an additional factor of ${\cal{B}}(\phi \to K^+ K^-)$, which is the branching fraction for the $\phi \to K^+ K^-$ decay, is included in the denominator of Eq.~\ref{eq:ul1}.

\begin{figure}[t]
\centering
\includegraphics[width=1.0\textwidth]{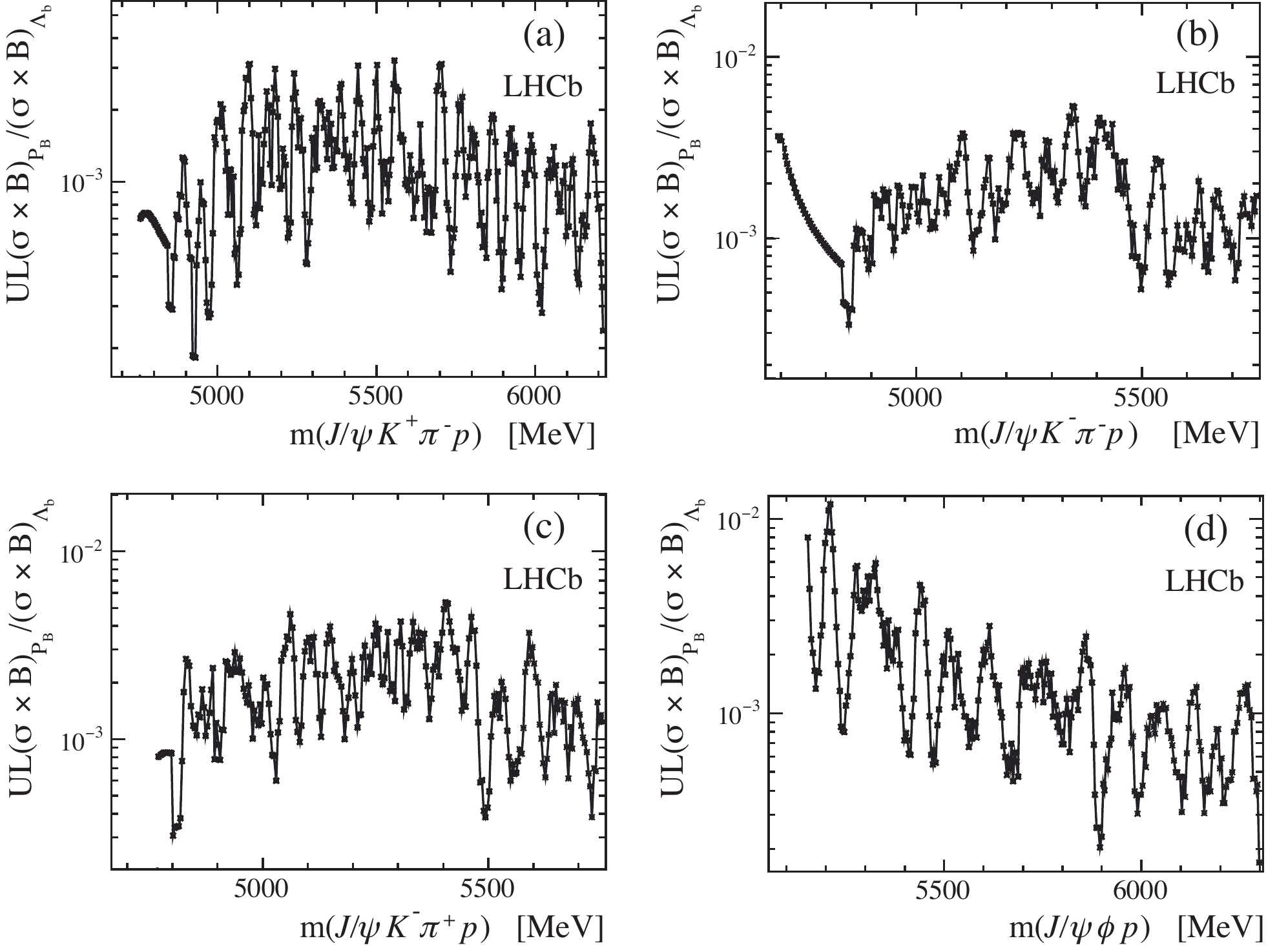}
\vskip -0.1cm
\caption{\small Upper limits on $R$ at 90\% {\rm CL}  for  (a)  $\jpsi K^+\pi^- p$, (b) $\jpsi K^-\pi^- p$, (c) $\jpsi K^-\pi^+ p$, and (d) $\jpsi \phi p$ final states.}
\label{ulnor}
\end{figure}

The systematic uncertainty on UL$(R)$ arises from the differences in analysis requirements between the search mode and the normalization mode (2\%), which is estimated based on the differences the selection requirements could make in the relative efficiencies. The detection of an additional track (1\%), given by the uncertainty in the data-driven tracking efficiency corrections,  and the identification of this track (1\%), given by the uncertainties in the particle identification calibration procedure, leads to an overall systematic uncertainty of 2.4\%. For mode \textbf{IV}, the small uncertainty on ${\cal{B}}(\phi \to K^+K^-$) is also taken into account. These uncertainties are added in quadrature with the uncertainty on $NF$. The upper limits on $R$ are then increased linearly by this small systematic uncertainty.
The results for UL($R$) at 90\% confidence level (CL) are shown in Fig.~\ref{ulnor}. Low invariant mass cut-offs in each mode are imposed when the efficiency uncertainty becomes large.

\section{Conclusions}
We have searched for pentaquark states containing a $b$ quark that decay weakly via the $b\to c\overline{c}s$ transition in the final states $\jpsi K^+\pi^- p$, $\jpsi K^- \pi^- p$, $\jpsi K^- \pi^+ p$, and $\jpsi \phi p$. Such states have been speculated to exist \cite{Klebanov,Stewart:2004pd,Leibovich:2003tw,Oh:1994np}. No evidence for these decays is found. Upper limits at 90\% confidence level on the ratio of the production cross sections of these states times the branching fractions into the search modes, with respect to the production and decay of the \Lb baryon in the mode $\jpsi K^- p$ ($R$, see Eq.~\ref{eq:R}) are found to be about $10^{-3}$, depending on the final state and the hypothesized mass of the pentaquark state. 
\section*{Acknowledgements}
%
%
\noindent We thank I. Klebanov for useful discussions. We express our gratitude to our colleagues in the CERN
accelerator departments for the excellent performance of the LHC. We
thank the technical and administrative staff at the LHCb
institutes. We acknowledge support from CERN and from the national
agencies: CAPES, CNPq, FAPERJ and FINEP (Brazil); MOST and NSFC
(China); CNRS/IN2P3 (France); BMBF, DFG and MPG (Germany); INFN
(Italy); NWO (The Netherlands); MNiSW and NCN (Poland); MEN/IFA
(Romania); MinES and FASO (Russia); MinECo (Spain); SNSF and SER
(Switzerland); NASU (Ukraine); STFC (United Kingdom); NSF (USA).  We
acknowledge the computing resources that are provided by CERN, IN2P3
(France), KIT and DESY (Germany), INFN (Italy), SURF (The
Netherlands), PIC (Spain), GridPP (United Kingdom), RRCKI and Yandex
LLC (Russia), CSCS (Switzerland), IFIN-HH (Romania), CBPF (Brazil),
PL-GRID (Poland) and OSC (USA). We are indebted to the communities
behind the multiple open-source software packages on which we depend.
Individual groups or members have received support from AvH Foundation
(Germany), EPLANET, Marie Sk\l{}odowska-Curie Actions and ERC
(European Union), ANR, Labex P2IO, ENIGMASS and OCEVU, and R\'{e}gion
Auvergne-Rh\^{o}ne-Alpes (France), RFBR and Yandex LLC (Russia), GVA,
XuntaGal and GENCAT (Spain), Herchel Smith Fund, the Royal Society,
the English-Speaking Union and the Leverhulme Trust (United Kingdom).

\afterpage{\clearpage}
\newpage
\bibliographystyle{LHCb}
\bibliography{b-pentaquark,LHCb-PAPER,LHCb-DP}
\ifx\mcitethebibliography\mciteundefinedmacro
\PackageError{LHCb.bst}{mciteplus.sty has not been loaded}
{This bibstyle requires the use of the mciteplus package.}\fi
\providecommand{\href}[2]{#2}

\newpage


 \centerline{\large\bf LHCb collaboration}
\begin{flushleft}
\small
R.~Aaij$^{40}$,
B.~Adeva$^{39}$,
M.~Adinolfi$^{48}$,
Z.~Ajaltouni$^{5}$,
S.~Akar$^{59}$,
J.~Albrecht$^{10}$,
F.~Alessio$^{40}$,
M.~Alexander$^{53}$,
A.~Alfonso~Albero$^{38}$,
S.~Ali$^{43}$
% ...

{\footnotesize \it
% ...
$ ^{5}$Clermont Universit{\'e}, Universit{\'e} Blaise Pascal, CNRS/IN2P3, LPC, Clermont-Ferrand, France\\
% ...
$ ^{10}$Fakult{\"a}t Physik, Technische Universit{\"a}t Dortmund, Dortmund, Germany\\
% ...
$ ^{38}$ICCUB, Universitat de Barcelona, Barcelona, Spain\\
$ ^{39}$Instituto Galego de F{\'\i}sica de Altas Enerx{\'\i}as (IGFAE), Universidade de Santiago de Compostela, Santiago de Compostela, Spain\\
$ ^{40}$European Organization for Nuclear Research (CERN), Geneva, Switzerland\\
% ...
$ ^{43}$Nikhef National Institute for Subatomic Physics, Amsterdam, The Netherlands\\
% ...
$ ^{48}$H.H. Wills Physics Laboratory, University of Bristol, Bristol, United Kingdom\\
% ...
$ ^{53}$School of Physics and Astronomy, University of Glasgow, Glasgow, United Kingdom\\
% ...
$ ^{59}$University of Cincinnati, Cincinnati, OH, United States\\
% ...
}
\end{flushleft}

\end{document}